\documentclass[12pt]{article}
\usepackage{amssymb,amsmath,epsfig,color}
\usepackage[colorlinks=true, pdfstartview=FitV, linkcolor=blue, citecolor=red, urlcolor=magenta]{hyperref}
\allowdisplaybreaks
\begin{document}
\title{\textbf{Slowly Rotating Wormholes in $f(R,T)$ Gravity:
Particle Dynamics and Weak-field Lensing}}
\author{M. Sharif$^{1,2}$\thanks{msharif.math@pu.edu.pk} ,
Tayyab Naseer$^{1}$\thanks{tayyabnaseer48@yahoo.com}~ and Rida
Batool$^{1}$\thanks{ridabatool.math@gmail.com}\\
$^1$Department of Mathematics and Statistics, The University of Lahore,\\
1-KM Defence Road Lahore-54000, Pakistan.\\
$^2$Research Center of Astrophysics and Cosmology, Khazar University, \\
Baku, AZ1096, 41 Mehseti Street, Azerbaijan.}
\date{}
\maketitle

\begin{abstract}
Among all gravitational solutions, traversable wormholes stand out
as truly fascinating, yet within general relativity, they almost
always demand exotic matter that necessarily violate null energy
requirement. The inclusion of matter-geometry coupling in modified
gravity theories provides a viable platform where wormholes remain
open which are supported by effective gravitational field. We
therefore examine the structure and properties of slowly rotating
traversable wormholes using the $f(R,T)$ framework. We formulate
stationary wormhole models coupled with the anisotropic fluid and
show that they are regular, horizon-free, approach a flat spacetime
geometry at large distances, and satisfy the flare-out requirement
at the throat. Our results indicate that the required matter meets
both the null and strong energy conditions, demonstrating that
rotating wormholes do not necessarily require unconventional
sources. We further study particle trajectories, frame dragging
phenomena, and non-geodesic behavior resulting from matter-geometry
coupling, alongside the weak-field and lensing effects caused by the
rotation. Our results establish the potential astrophysical
viability of rotating wormholes in $f(R,T)$ gravity.
\end{abstract}

\section{Introduction}

Astronomical data increasingly support the idea that the universe is
expanding faster over time, prompting investigations into its
underlying mechanism \cite{81,82}. Observations of type la
supernovae and other astronomical measurements provide strong
evidence that the cosmic expansion is occurring at an accelerating
rate, challenging earlier cosmological expectations \cite{1,1a}. The
accelerated growth of the universe is commonly attributed to dark
energy, a component thought to exert a repulsive influence.
Scientists have therefore explored various theoretical frameworks to
interpret the observational evidence. Modern cosmological research
largely relies on the $\Lambda$-CDM model, the standard framework
for studying dark energy and the universe' evolution. These
observations clarify the impact of dark energy on the geometry and
large-scale organization of the cosmos. While highly successful in
explaining major cosmological features, two unresolved theoretical
concerns continue to affect the $\Lambda$-CDM model: why cosmic
densities are comparable today and why its parameters require such
precise adjustment. Resolving these issues could demand a deeper
understanding of gravity, potentially leading to alternative
theoretical models.

Buchdahl proposed $f(R)$ gravity as an extension of general
relativity (GR) by replacing $R$ (the Ricci scalar) in the
Einstein-Hilbert action with a differentiable function $f(R)$
\cite{66}. This theoretical framework has been widely used to
analyze the characteristics and progression of different
astrophysical phenomena \cite{2}-\cite{4}. Recent evidence from
cosmology and gravitational physics has raised the possibility that
GR may not completely describe all observed phenomena, supporting
interest in $f(R)$ gravity. Through an extensive analysis, Bamba et
al. \cite{90b} explored different dark energy models within a
$\Lambda$-CDM type cosmological framework. In \cite{5a}, Nojiri and
Odintsov provided a detailed theoretical analysis of $f(R)$ gravity.
They demonstrated its consistency with both cosmological
observations and local tests, as well as its relevance to
gravitational baryogenesis. In a recent study, Agrawal et al.
\cite{90c} analyzed the $f(R)$ cosmological model containing Ricci
scalar contributions and a homogeneous perfect fluid. Bertolami et
al. \cite{90d} introduced the $f(R,L_m)$ framework as an extension
of gravity that directly couples the matter Lagrangian $L_m$ with
spacetime geometry. By explicitly introducing the matter Lagrangian,
the $f(R)$ framework allows researchers to explore its influence on
different spacetime geometries.

Harko et al. \cite{67} extended $f(R)$ gravity to $f(R,T)$ by
incorporating the trace $T$ of energy-momentum tensor (EMT). In this
framework, the gravitational field equations are derived by varying
the action with respect to the metric. The non-conservation of the
EMT gives an extra force whose magnitude depends on thermodynamic
variables such as density and pressure \cite{86}. This feature may
have important implications for the evolution of astrophysical
systems. The additional force causes particles to follow
non-geodesic trajectories, representing a notable departure from GR.
Singh and Kumar \cite{88} considered how the trace term influences
cosmic expansion when the coupling constant is assigned distinct
values. Sharif and Zubair \cite{89} further showed that the model
can describe both phantom and non-phantom expansion phases while
remaining broadly consistent with $\Lambda$-CDM evolution. Baffou et
al. \cite{87} assessed the observational viability of the two
cosmological models. Their results satisfied observational
constraints over a broad redshift range, covering both nearby and
distant cosmic sources. Tretyakov \cite{77} investigated the
stability conditions of the theory after including matter terms with
higher-order derivatives, with particular attention to ghosts,
tachyons, and related instabilities. This gravitational framework
has been widely used to the study of various astrophysical and
cosmological phenomena \cite{77a}-\cite{90a}.

The properties and evolution of wormholes (WHs) are being
investigated through astronomical observations and theoretical
models. Such studies may help determine whether these objects can
serve as hypothetical pathways across spacetime. A WH throat is its
smallest cross-sectional surface, serving as a connection between
two different regions of spacetime. By analogy, it resembles a metro
tunnel that creates a direct link between far-apart urban locations.
A key feature of a WH is two way traversability, which depends on
the stability of its throat. Traversable WHs can therefore act as
highly efficient shortcuts through spacetime. Wormholes are
generally classified into two main categories: \emph{(i)}
intra-universe WHs, which join distant regions of the same universe,
and \emph{(ii)} inter-universe WHs, which connect distinct
universes. This indicates that WHs may provide connections to other
universes, each of which could obey its own physical laws. A WH is
termed static if its throat radius remains constant over time. In
contrast to static WHs, whose throat radius remains constant,
dynamic WHs have a time dependent throat that changes the geometry
of the spacetime passage. Classifying these astrophysical phenomena
can support the development of theoretical models for traversable
WHs and non-local communication.

Wheeler's pioneering contributions \cite{84,84a}, followed by his
work with Misner \cite{83,83a}, played a central role in
establishing theoretical WH models. Their work interpreted
electromagnetic charge in terms of multiply connected spacetime
configurations that allow field lines to traverse non-trivial
topological paths. Einstein and Rosen \cite{16a} provided a detailed
analysis of the framework through the Einstein-Rosen bridge. As a
precise solution of Einstein field equations, it offered one of the
earliest geometric models of a throat connecting distant parts of
spacetime. According to Fuller and Wheeler \cite{17a}, such
geometrical structures are intrinsically unstable under physical
perturbations. They found that the bridge rapidly collapses after
formation and therefore cannot transmit photons. The effect can be
attributed to the close relationship between non-trivial spacetime
geometry and gravitational dynamics in regions of strong curvature.
According to Fuller and Wheeler \cite{85}, the Schwarzschild WH is
non-traversable because its geometry does not fulfill the essential
conditions required for light to pass through it.

The rotation of compact astrophysical objects plays a significant
role in their physical behavior. For WHs, it can introduce frame
dragging, modify particle orbits, and produce observational
features, which cannot be found in static configurations \cite{19a}.
Rotation has a significant influence on the system's dynamics of
massive particles and light rays, with consequences for the
deflection of light by gravity, stable trajectories, and
gravitational lensing. Exact solutions for rotating WHs are
typically challenging to derive analytically. For this reason, the
slow rotation approximation is widely used to investigate the
primary effects of rotation while preserving a manageable analytical
description of the geometry. These considerations lead us to examine
rotating traversable WH solutions in $f(R,T)$ gravity. Extending
previous studies of static WHs in $f(R,T)$ gravity
\cite{aa}-\cite{ll}, we investigate rotational effects on particle
motion, frame dragging, and weak-field lensing. We also carry out a
detailed examination of the additional force generated by
interaction of matter with spacetime geometry, and the corresponding
non-geodesic corrections, thereby improving our knowledge of
rotating WHs' dynamical and observational behavior in modified
gravity theory. Our results imply that rotating WH geometries can be
physically maintained by matter-geometry interaction within $f(R,T)$
gravity, and they may display characteristics that set them apart
from analogous GR configurations.

This is how we have organized the paper. Section \textbf{2}
introduces the $f(R,T)$ gravity formalism and derives the
non-vanishing components of the field equations. In section
\textbf{3}, the rotating WH solution is constructed, followed by an
analysis of the shape functions and rotational profile. We also
examine the absence of exotic fluid by analyzing the relevant energy
conditions. The solutions are further examined in section \textbf{4}
for their geometrical properties and stability. In section
\textbf{5}, we investigate particle dynamics, frame dragging, the
additional force produced by matter-geometry coupling, and
weak-field and gravitational lensing. Finally, section \textbf{6}
presents the concluding remarks, consolidating the central results
and emphasizing the potential of $f(R,T)$ gravity to support
physically realistic, non-exotic rotating WH configurations with
observable astrophysical signatures.

\section{Fluid-geometry Coupling and Field Equations}

In the context of $f(R,T)$ modified gravity, the corresponding
action reads
\begin{equation}\label{1}
I=\int \sqrt{-g}\bigg[\frac{f(R,T)}{16\pi}+L_m\bigg] d^{4}x.
\end{equation}
Taking $g$ to be the determinant associated with the metric, we let
$L_m$ represent the matter Lagrangian density. In this alternative
gravitational description, the generalized form of the field
equations is
\begin{equation}\label{2}
G_{\alpha\xi}=8 \pi T_{\alpha\xi}^{(tot)}.
\end{equation}
In this relation, the Einstein tensor $G_{\alpha\xi}$ accounts for
the geometrical structure of spacetime and the quantity
$T_{\alpha\xi}^{\text{(tot)}}$ gives the total EMT arising from the
cosmic matter distribution. Its definition is presented as
\begin{equation}\label{3}
T_{\alpha\xi}^{(tot)}=
\frac{1}{f_R}\big\{T_{\alpha\xi}+T_{\alpha\xi}^{(c)}\big\}.
\end{equation}
The quantities $T_{\alpha\xi}$ and $T_{\alpha\xi}^{(c)}$ are
identified with the ordinary matter component and the supplementary
effects due to curvature-matter coupling within this modified
gravity, respectively. The final term of Eq.\eqref{3} takes the
following form
\begin{align}\nonumber
T_{\alpha\xi}^{(c)}&=\frac{1}{8\pi
f_R}\bigg[2f_TT_{\alpha\xi}+\bigg\{\frac{1}{2}(f-Rf_R)-f_TL_m\bigg\}g_{\alpha\xi}
\\\label{4}
&-(g_{\alpha\xi}\Box-\nabla_{\alpha}\nabla_{\xi})f_{_{R}}+2 f_T
g^{\nu\chi}\frac{\partial^2 L_m}{\partial g^{\alpha\xi}\partial
g^{\nu\chi}}\bigg],
\end{align}
where the subscripts $R$ and $T$ on $f$ indicate differentiation
with respect to the corresponding terms. The D'Alembertian operator
is subsequently expressed in terms of
$\Box\equiv\frac{1}{\sqrt-g}\partial_{\alpha}(\sqrt-g
g^{\alpha\xi}\partial_{\xi})$) along with the covariant derivative
$\nabla_{\xi}$. From a combination of the above equations, we obtain
\begin{align}\nonumber
G_{\alpha\xi}&=\frac{1}{f_R}\bigg[(8\pi+f_T)T_{\alpha\xi}
-\big(g_{\alpha\xi}\Box-\nabla_{\alpha}\nabla_{\xi}\big)f_{_{R}}\\\label{5}
&+\frac{1}{2}g_{\alpha\xi}(f-Rf_R)-f_Tg_{\alpha\xi}L_m+2 f_T
g^{\nu\chi}\frac{\partial^2 L_m}{\partial g^{\alpha\xi}\partial
g^{\nu\chi}}\bigg].
\end{align}

The identification of numerous physical processes giving rise to
pressure anisotropy has significantly expanded research into stellar
fluid configurations within astrophysical contexts. In essence, the
pressure anisotropy observed in compact stars is driven by the
cumulative effect of several physical mechanisms operating together.
Turning to our WH scenario, the anisotropic fluid is set up as
follows
\begin{equation}\label{6}
T_{\alpha\xi}=(\rho+p_{t})v_{\alpha}v_{\xi}
+(p_{r}-p_{t})k_{\alpha}k_{\xi}+p_{t}g_{\alpha\xi}.
\end{equation}
To examine the geometry of the WH, we first lay down the physical
requirements that ensure a stable and traversable throat. Here, the
anisotropy of the fluid is defined as the difference $p_t - p_r$,
where $p_r$ and $p_t$ are the radial and transverse pressures, and
$\rho$ is the local energy density. In this formulation, $v_\alpha$
stands for the four-velocity, and $k_\alpha$ signifies a spacelike
four-vector.

The modified gravity theory based on the functional $f(R,T)$ gives
rise to several possible models, each with its own matter-geometry
coupling intensity ranging from minimal to non-minimal. Our
investigation utilizes a model that is both physically meaningful
and mathematically workable, written as
\begin{equation}\label{7}
f(R,T)=f_1(R)+f_2(T)=R+\lambda T,
\end{equation}
where $\lambda$ is taken as a real-valued parameter. The special
relevance of this model stems from its prospective resolution of the
age-old cosmological constant issue \cite{mm,nn}. According to
recent work in scale-invariant gravity and dynamical vacuum
cosmology, a cosmological constant that evolves over time may
provide the key to solving these problems \cite{oo,pp}.

\subsection{Wormhole Geometry under Slow Rotation}

We consider a stationary spacetime to investigate rotating
traversable WHs in $f(R,T)$ gravity. The static Morris-Thorne WH is
generalized in such geometries via the inclusion of an off-diagonal
metric element that accounts for frame dragging. The line element
for this rotating case can be cast in the form
\begin{align}\label{8}
ds^{2}&=-N^{2}(r)dt^{2}+\frac{dr^{2}}{1-\frac{b(r)}{r}}+r^{2}\{d\theta^{2}
+\sin^{2}\theta(d\phi-\omega(r)dt)^{2}\}.
\end{align}
The terms $N(r)$ and $b(r)$ describe the redshift and shape
functions, respectively, while $\omega(r)$ represents the frame
dragging angular velocity. For the WH to stay horizonless and
traversable, the redshift function needs to be finite and non-zero
at all points. The shape function, which defines the spatial
configuration, is additionally constrained by the throat condition
\begin{align}\label{9}
b(r_0)&=r_0.
\end{align}
Here, $r_0$ represents the WH throat radius. For a physically viable
WH geometry, the flare-out condition must hold at or close to the
central neck \cite{tt}
\begin{align}\label{10}
\frac{b(r)-rb'(r)}{b^{2}(r)}>0.
\end{align}
Moreover, asymptotic flatness requires the rotational function to
vanish at spatial infinity
\begin{align}\label{11}
\lim_{r\rightarrow \infty}\omega(r)=0.
\end{align}

As exact rotating WH solutions are difficult to obtain in closed
form, we employ the slow rotation approximation and retain
$\omega(r)$ up to the linear order
\begin{align}\label{12}
\omega(r)&=\epsilon\,\Omega(r),\quad \epsilon\ll1.
\end{align}
Here, $\epsilon$ plays the role of a dimensionless perturbative
parameter. By expanding the metric to linear order in $\epsilon$, it
follows that
\begin{align}\label{13}
\sin^{2}\theta\,(d\phi-\omega(r)\,dt)^{2} &=
\sin^{2}\theta\,d\phi^{2} -2\,\omega(r)\,\sin^{2}\theta\,dt\,d\phi
+\mathcal{O}(\epsilon^{2}),
\end{align}
and consequently, we obtain the metric in the form
\begin{align}\label{14}
ds^{2} &\simeq -N^{2}(r)\,dt^{2} +\frac{dr^{2}}{1-\dfrac{b(r)}{r}}
+r^{2}d\theta^{2} +r^{2}\sin^{2}\theta\,d\phi^{2}
-2r^{2}\sin^{2}\theta\,\omega(r)\,dt\,d\phi.
\end{align}
Under this approximation, the leading order frame dragging effects
are retained separately, while the zeroth-order metric remains
exactly that of the static Morris-Thorne WH, i.e., When
$\omega\rightarrow0$, Eq.\eqref{8} describes the usual static
traversable WH spacetime. We also define
\begin{align}\label{15}
\Delta(r)&\equiv 1-\frac{b(r)}{r}.
\end{align}

For metric functions depending solely on $r$, substitution of
Eq.\eqref{8} into the modified field equation \eqref{5} leads to the
radial behavior of $N(r)$, $b(r)$, and $\omega(r)$. For the linear
model \eqref{7}, metric \eqref{14}, and the choice \eqref{15}, the
field equations \eqref{5} takes the form
\begin{align}\nonumber
&N(r)\bigg[ \Delta(r)N''(r) +\bigg( \frac{2}{r} -\frac{b'(r)}{2r}
-\frac{3b(r)}{2r^{2}} \bigg)N'(r) \bigg] +\frac{1}{2} \bigg[ \bigg\{
\frac{2\Delta(r)N''(r)}{N(r)} \\\nonumber & +\frac{2}{r} \bigg(
2-\frac{rb'(r)+3b(r)}{2r} \bigg) \frac{N'(r)}{N(r)}
+\frac{2b'(r)}{r^{2}} \bigg\} +\lambda\big(
-\rho(r)+p_r(r)\\\label{16}&+2p_t(r) \big) \bigg] N^{2}(r) = 8\pi
\rho(r)N^{2}(r),
\\\nonumber &\bigg[ -\frac{N''(r)}{N(r)}
+\frac{rb'(r)-b(r)}{2r\big(r-b(r)\big)} \frac{N'(r)}{N(r)}
+\frac{rb'(r)-b(r)}{r^{2}\big(r-b(r)\big)} \bigg]
-\frac{1}{2\Delta(r)}\\\nonumber & \times \bigg[
\frac{2\Delta(r)N''(r)}{N(r)} +\frac{2}{r} \bigg\{
2-\frac{rb'(r)+3b(r)}{2r} \bigg\} \frac{N'(r)}{N(r)}
+\frac{2b'(r)}{r^{2}}\\\label{17}&
+\lambda\big(-\rho(r)+p_r(r)+2p_t(r)\big) \bigg] =
\frac{1}{\Delta(r)}\{{(8\pi+\lambda)} p_r(r) -\lambda \rho(r)\},
\\\nonumber &\bigg[ b'(r) +\big(b(r)-r\big)\frac{N'(r)}{N(r)} -\frac{b(r)-rb'(r)}{2r} \bigg]
+\frac{1}{2} \bigg[ \bigg\{ \frac{2\Delta(r)N''(r)}{N(r)}
\\\nonumber &-\frac{2}{r} \bigg( 2-\frac{rb'(r)+3b(r)}{2r} \bigg)
\frac{N'(r)}{N(r)} -\frac{2b'(r)}{r^{2}} \bigg\}
+\lambda\big(-\rho+p_r+2p_t\big)r^{2} \bigg]\\\label{18}& =
(8\pi+\lambda) p_t(r) r^{2} + r^2\lambda\rho(r),
\\\nonumber &-\frac{r}{2}\Delta\omega''
- \frac{1}{2}\bigg(4 - \frac{rb'+3b}{2r} +
\frac{2rN'}{N}\bigg)\Delta\omega' + r^2\bigg[\frac{\Delta N''}{N} -
\frac{(4r-rb'-3b)}{2r^2}\frac{N'}{N} \\\label{19} &-
\frac{b'}{r^2}\bigg] =- \lambda r^2 p_r(r)-p_t(r)\{ 8\pi r^2
+3\lambda\} .
\end{align}
Solving these equations simultaneously gives the full mathematical
form of matter variables as
\begin{align}\nonumber
\rho&=-\frac{1}{8 \left(-\lambda ^3+3 \pi  \lambda ^2-24 \pi ^2
\lambda +128 \pi ^3\right) r^4 (r-b(r))}\big[-2 \lambda ^2 r^4
b'(r)-16 \pi \lambda r^4 b'(r)\\\nonumber &+3 \lambda ^2 r^3
b'(r)+40 \pi \lambda r^3 b'(r)-128 \pi ^2 r^3 b'(r)-\lambda ^2 r^2
b(r) b'(r)-24\pi \lambda r^2 b(r) b'(r)\\\nonumber &+128 \pi ^2 r^2
b(r) b'(r)+4 \lambda ^2 r b'(r)-4 \lambda ^2 b(r) b'(r)-32 \pi
\lambda  r b'(r)+32 \pi \lambda b(r) b'(r)\\\label{20} &+2 \lambda
^2 r^3 b(r)+16 \pi \lambda r^3 b(r)+2 \lambda ^2 r^2 b(r)-16 \pi
\lambda r^2 b(r)-4 \lambda ^2 r b(r)^2\big],
\\\nonumber p_r&=\frac{1}{8 \big(-\lambda ^3+3 \pi  \lambda ^2-24 \pi ^2 \lambda
+128 \pi ^3\big) r^4 (r-b(r))}\big[6 \lambda ^2 r^4 b'(r)+128 \pi ^2
r^4 b'(r)\\\nonumber&-9 \lambda ^2 r^3 b'(r)+16 \pi \lambda r^3
b'(r)-128 \pi ^2 r^3 b'(r)+3 \lambda ^2 r^2 b(r) b'(r)-16 \pi
\lambda r^2 b(r) b'(r)\\\nonumber&+4 \lambda ^2 r b'(r)-4 \lambda ^2
b(r) b'(r)+16 \pi \lambda r b'(r)-16 \pi \lambda b(r) b'(r)-6
\lambda ^2 r^3 b(r)\\\label{21}&-128 \pi ^2 r^3 b(r)+2 \lambda ^2
r^2 b(r)+8 \pi \lambda r^2 b(r)+4 \lambda ^2 r b(r)^2-8 \pi \lambda
r b(r)^2+128 \pi ^2 r b(r)^2\big],
\\\nonumber p_t&=-\frac{1}{8 \big(-\lambda ^3+3 \pi \lambda ^2-24 \pi ^2 \lambda
+128 \pi^3\big) r^4 (r-b(r))}\big[2 \lambda ^2 r^4 b'(r)-16 \pi
\lambda r^4 b'(r)\\\nonumber&-7 \lambda ^2 r^3 b'(r)+52 \pi \lambda
r^3 b'(r)-64 \pi ^2 r^3 b'(r)+5 \lambda ^2 r^2 b(r) b'(r)-36 \pi
\lambda r^2 b(r) b'(r)\\\nonumber&+64 \pi ^2 r^2 b(r) b'(r)+4
\lambda ^2 r b'(r)-4 \lambda ^2 b(r) b'(r)+8 \pi \lambda  r b'(r)-8
\pi \lambda b(r) b'(r)\\\nonumber&-128 \pi ^2 r b'(r)+128 \pi ^2
b(r) b'(r)-2 \lambda ^2 r^3 b(r)+16 \pi \lambda r^3 b(r)+2 \lambda
^2 r^2 b(r)\\\label{22}&+4 \pi  \lambda r^2 b(r)-64 \pi ^2 r^2
b(r)-20 \pi \lambda  r b(r)^2+64 \pi ^2 r b(r)^2\big].
\end{align}
Collectively, these equations govern the structure of slowly
rotating traversable WHs within the linear $f(R,T)$ gravity.

\section{Wormhole Solutions with Rotation}

Having established the governing equations, we now move forward to
obtain explicit rotating WH solutions within the present formalism.
With suitable shape and rotational functions, traversable and
asymptotically-flat geometries can be obtained within the modified
framework. Next, we turn to the matter distribution and the role of
matter-geometry coupling in determining the physical viability of
the WH solutions.

\subsection{Shape Function and Rotational Profile}

The reconstruction method is employed to generate explicit
solutions. The analysis begins with appropriate WH metric functions
satisfying the necessary geometric conditions, from which the matter
sector is derived through the modified field equations. This
reconstruction technique is commonly employed in the WH literature,
as it allows one to fix the metric functions to capture the desired
topological features and then determine the matter distribution that
ensures consistency with the gravitational field equations. Moving
forward, we choose to work with a constant redshift function, given
by
\begin{align}\label{23}
N(r)=1,
\end{align}
which not only prevents the formation of horizons but also reduces
the complexity of the field equations, while preserving the key
properties of the WH. We adopt two distinct choices for the shape
function
\begin{itemize}
\item
The first choice (considered as model 1) is given by \cite{uuu}
\begin{align}\label{24}
b(r)=r_{0}\left(\frac{\cosh(r_{0})}{\cosh(r)}\right)^{\mu}, \quad
0<\mu<1,
\end{align}
where $\mu$ is a free parameter. The throat condition $b(r_0)= r_0$
is automatically satisfied, and differentiation gives $b'( r_0 ) =
-\mu \tanh(r_0)$. Hence, as long as $\mu \tanh(r_0) < 1$, one has
$b'(r_0) < 1$ , and the flare-out requirement holds at the throat.
\item
The second choice (considered as model 2) takes the form \cite{uu}
\begin{align}\label{25}
b(r)=\frac{r_{0}\,a^{r}}{a^{r_{0}}}, \quad a\in(0,1),
\end{align}
where  $a\in(0,1)$ is a constant. This form satisfies $b(r_0) =
r_0$, and gives $b'(r_0)= ln(a)<0$. Since $ln(a)<0$  for
$a\in(0,1)$, we automatically have $b'(r_0)< 1$, thus the flare-out
requirement is fully met at the throat for every allowed $a$.
\end{itemize}

Once we fix the rotation profile from the $(t\phi)$ part of the
modified field equations \eqref{19}, and with slow rotation plus no
off-diagonal matter source, it becomes much simpler as
\begin{align}\label{26}
\left(r-b(r)\right) \omega''(r) +\left( \frac{4r-3b(r)-rb'(r)}{r}
\right)\omega'(r) =0.
\end{align}
In terms of the shape function, Eq.\eqref{26} gives
\begin{align}\label{27}
\omega(r) = C_{1} + C_{2} \int \frac{dr}
{r^{4}\sqrt{1-\dfrac{b(r)}{r}}},
\end{align}
where $C_1$ and $C_2$ are constants of integration. Asymptotic
flatness requires $\omega(r)\rightarrow0$ as $r\rightarrow\infty$,
thereby fixing the constant part. In the large-$r$ limit,
$b(r)/r\rightarrow0$, so Eq.\eqref{27} simplifies to
\begin{align}\label{28}
\omega(r)\sim\frac{C}{r^{3}}, \quad r\rightarrow\infty.
\end{align}
Consequently, the rotational profile decays in the usual way under
frame dragging effects, in close analogy with the Lense-Thirring
phenomenon known for slowly rotating dense objects. This guarantees
that the effects of rotation become visible in the vicinity of the
central neck, while the standard flat asymptotic profile is regained
far away.
\begin{figure}[h!]
\centering
\epsfig{file=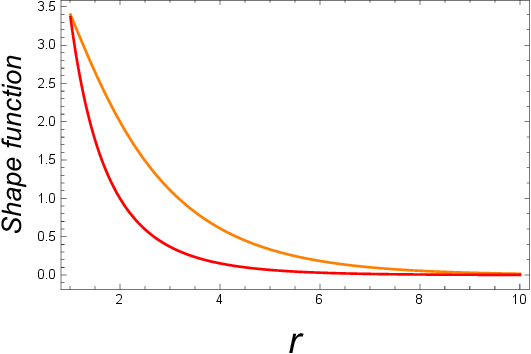,width=0.42\linewidth}\hfill\epsfig{file=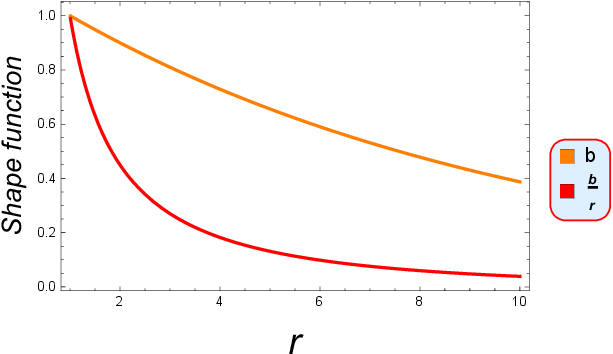,width=0.49\linewidth}
\caption{Shape function for model 1 \eqref{24} (left) and model 2
\eqref{25} (right).} \label{fig:ShapeFunction}
\end{figure}

The behavior of the shape functions for both WH models is presented
in Figure \textbf{1}. Each function satisfies $b(r_0)=r_0$ at the
throat and decreases with radial distance. The corresponding
flare-out condition $b'(r_0)<1$ is also satisfied, confirming the
suitability of both shape functions for constructing traversable WH
geometries. Model \textbf{1} exhibits a more rapid decrease of the
shape function with $r$, whereas model \textbf{2} decreases more
gradually.

\subsection{Profile of Energy Conditions}

Fundamental restrictions on EMT governing the fluid content in
spacetime are provided by energy conditions, which serve to ensure
that gravitational solutions are physically acceptable. These
criteria are regarded as essential for interpreting the relationship
between energy and matter. A variety of components dark energy,
exotic matter, baryonic matter, and dark matter are known to
constitute the universe, and each is endowed with its own physical
characteristics. One of the main theoretical challenges in
astrophysics is therefore the identification of spacetime geometries
that can satisfy all energy conditions.

Two fundamentally different possibilities are distinguished in WH
physics based on whether energy conditions are satisfied or
violated. When they are satisfied, WHs are permitted without exotic
matter; and otherwise, they often need exotic matter for a
physically viable solution. This analysis is based on the four
standard energy conditions: weak, dominant, strong, and null. A
comprehensive analysis of all energy conditions is undertaken in our
WH research. These constraints are taken as the foundational
requirements for physically admissible geometries. Mathematically,
they are stated as
\begin{equation}\nonumber
\rho > 0, \quad \rho + p_r > 0, \quad \rho + p_t > 0, \quad \rho +
p_r + 2p_t > 0.
\end{equation}

These conditions are investigated for each WH model using the
parametric values as $\lambda=15,20,25,30,35,40$, $a=0.4$,
$\mu=0.5$, and $L=0,2,4$. Careful tuning of the parameters that
regulate all energy conditions is required to maintain traversable
WH stability. When they are met, stable curvature and realistic
matter are obtained, and exotic matter is not required. Due to
intrinsic limitations in different gravity theories and the
parameter choices that are made, the construction of realistic WH
models without exotic matter may become harder. Additional
investigations are necessary if parameter selection for traversable
WHs is to be improved. Both proposed models are confirmed to be
physically feasible without exotic matter, highlighting the
framework's importance. Figure \textbf{2} shows that the plotted
energy conditions remain positive for both models. They decrease
rapidly with increasing $r$ and approach zero at large $r$,
indicating that the considered energy conditions are satisfied and
become less significant away from the throat.
\begin{figure}[h!]
\centering\epsfig{file=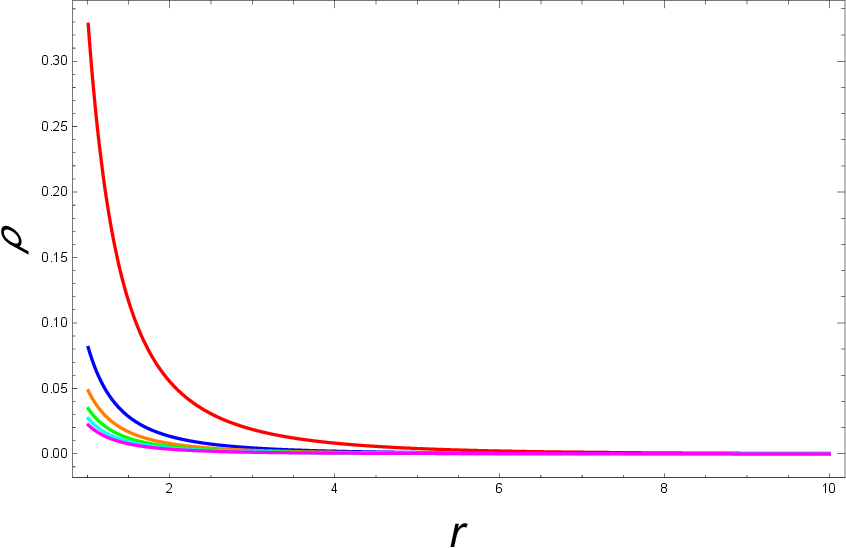,width=0.46\linewidth}\hfill\epsfig{file=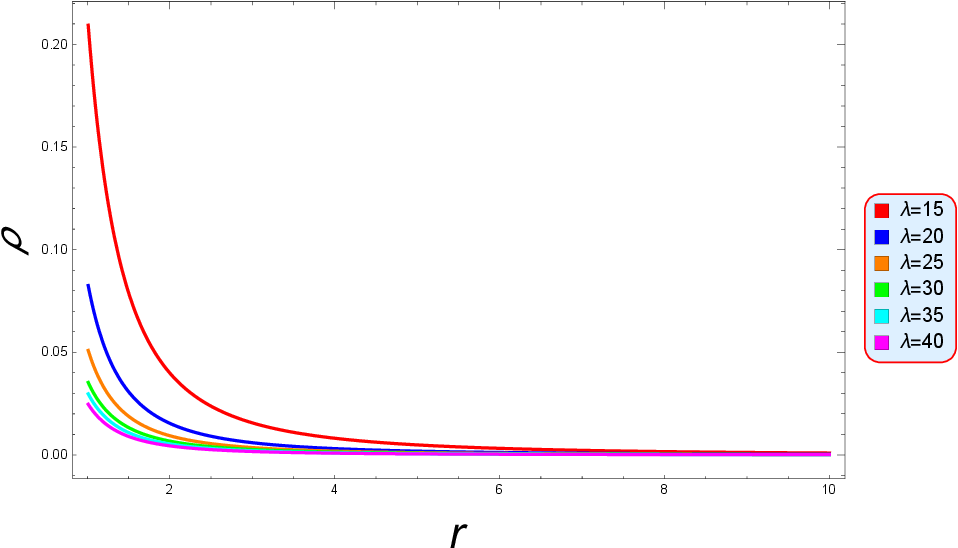,width=0.525\linewidth}
\epsfig{file=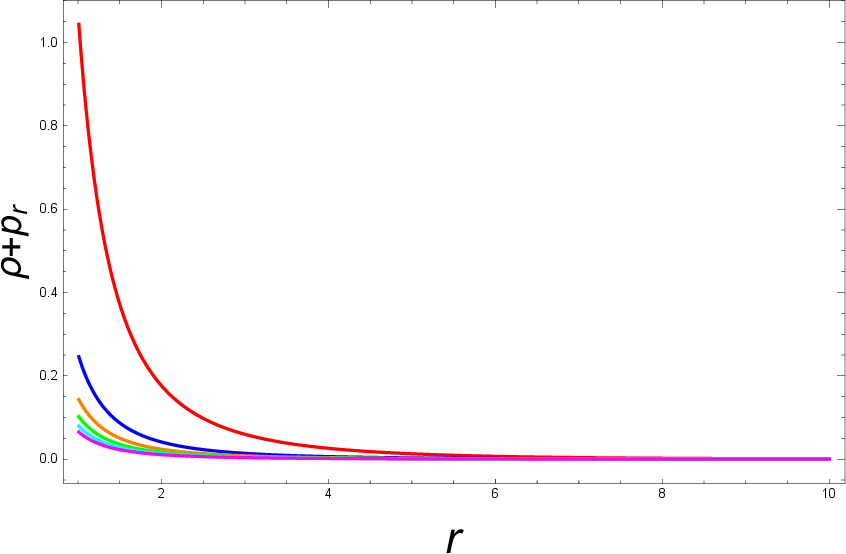,width=0.46\linewidth}\hfill\epsfig{file=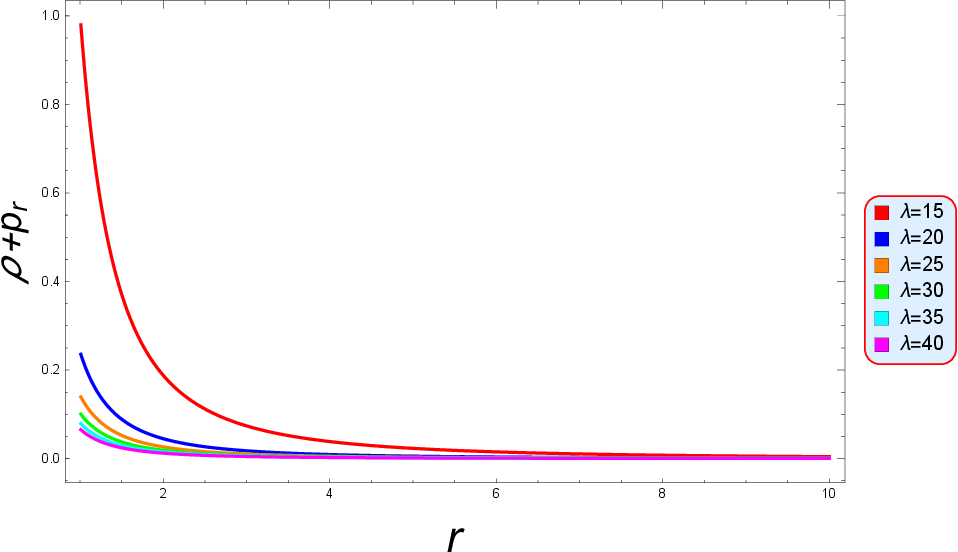,width=0.525\linewidth}
\epsfig{file=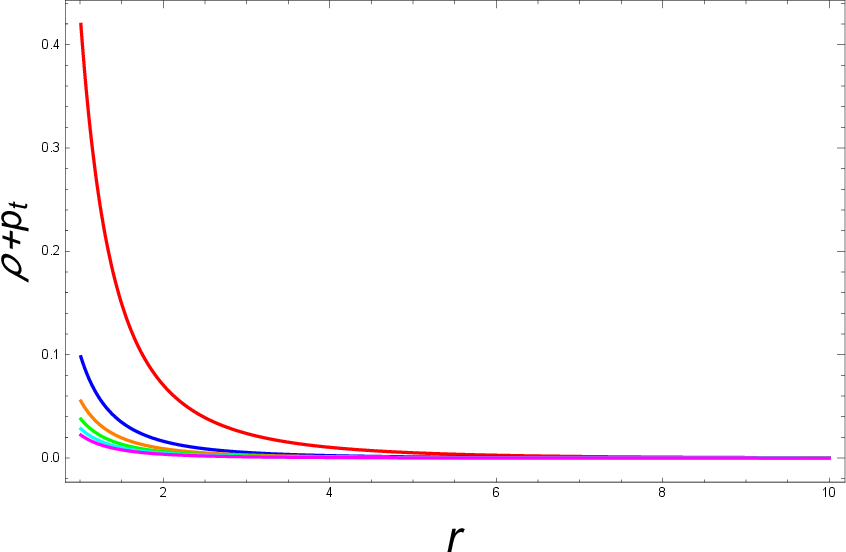,width=0.46\linewidth}\hfill\epsfig{file=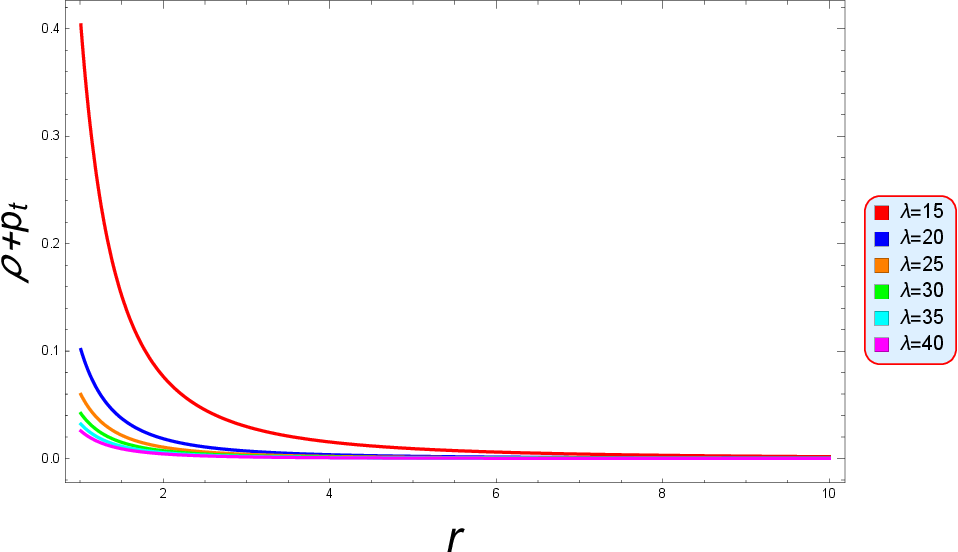,width=0.525\linewidth}
\epsfig{file=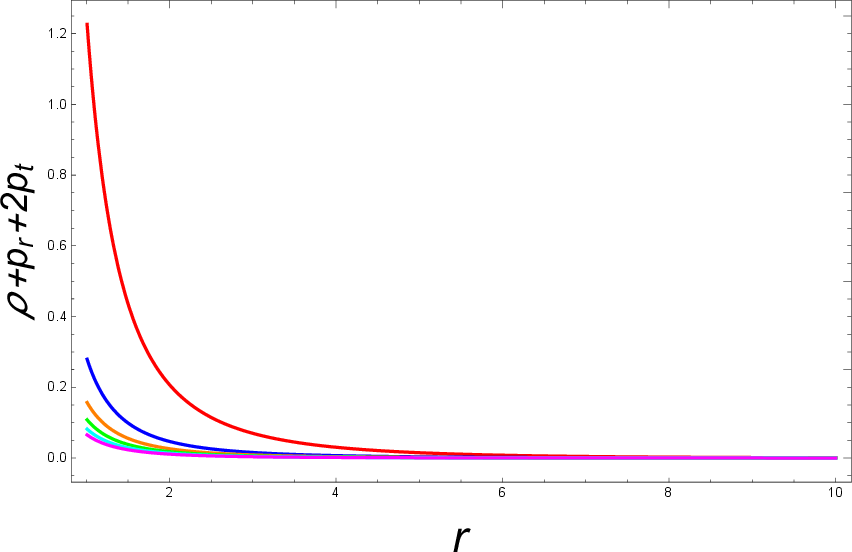,width=0.46\linewidth}\hfill\epsfig{file=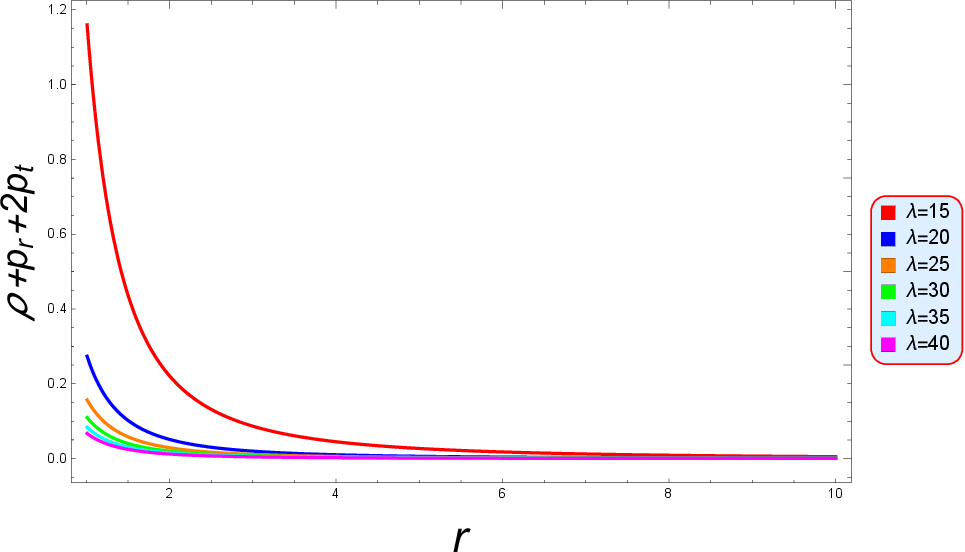,width=0.525\linewidth}
\caption{Energy conditions for model 1 (left) and model 2 (right).}
\label{fig:Energy conditions}
\end{figure}

\section{Geometrical Properties and Stability}

We now proceed to analyze the geometric structure and physical
consistency of the rotating WH solutions obtained above. We focus
specifically on the embedding structure of our solutions, using it
as a tool to visualize the WH geometry and to ascertain that the
spacetime is indeed traversable. We also examine whether the
solutions satisfy elementary stability requirements and remain
compatible with the matter-geometry coupled gravity setup.

\subsection{Embedding Diagram}

For the purpose of visualizing the WH's spatial geometry, we employ
an embedding diagram, which effectively maps the curvature and
topology of the spacetime. Taking advantage of spherical symmetry,
we can focus exclusively on the equatorial plane, which greatly
simplifies the analysis, which is determined by
$\theta=\frac{\pi}{2}$. This procedure keeps the necessary geometric
details intact while effectively reducing the dimensionality to two.
Fixing the temporal coordinate $t$ allows us to examine a particular
slice of the WH geometry. This yields a two-dimensional line element
that allows us to visualize the spatial structure and assess the
WH's connectivity. It is expressed as
\begin{align}\label{29}
ds^2= \frac{dr^2}{1-\dfrac{b(r)}{r}} +r^2d\phi^2.
\end{align}

Cylindrical coordinates $(r,\phi,z)$ are used to describe the
spatial curvature: $r$ is radial, $\phi$ is azimuthal, and $z$ is
axial. The axisymmetric geometry of the WH is clearly described
using this coordinate system, and the curvature is also made easier
to visualize. Proper comprehension of the WH geometry is achieved
through its mathematical embedding in a well-defined coordinate
system. Consequently, the distorted spacetime is represented as a
two-dimensional sheet embedded in a flat three-dimensional space.
For all WH configurations, the embedding geometry is uniquely
specified by Eq.\eqref{29}, and how it looks exactly depends on the
metric potentials used. This is given by
\begin{equation}\label{30}
ds^2=dr^2+r^2d\phi^2+dz^2.
\end{equation}

The axisymmetric structure of the WH is effectively represented by
an embedded surface defined as $z(r)$ within a three-dimensional
Cartesian coordinate system as
\begin{equation}\label{31}
ds^2=dr^2\bigg\{1+\bigg(\frac{dz}{dr}\bigg)^2\bigg\}+r^2d\phi^2.
\end{equation}
Equations \eqref{29} and \eqref{30} produce
\begin{equation}\label{32}
\frac{dz}{dr}=\pm \frac{1}{\sqrt{\frac{r}{b(r)}-1}}.
\end{equation}
A key geometric feature is seen at the throat: the embedded surface
becomes vertical, as shown by $\frac{dz}{dr} \to \infty $. This
behavior is associated with a strong constriction of the neck, which
is found to correspond the area of greatest curvature. The
asymptotic flatness condition is satisfied when the slope
$\frac{dz}{dr}$ equals zero as $r\rightarrow\infty$, i.e., the
surface is seen to become parallel to the radial axis at large $r$,
showing that the spacetime becomes flat. The following relation is
used to govern this asymptotic property mathematically
\begin{equation}\label{33}
z(r)=\pm \int_{r_{0}}^{r} \frac{dr}{\sqrt{\frac{r}{b(r)}-1}}.
\end{equation}

On either side of the throat, the spacetime is asymptotically flat.
So the embedding diagram is taken to represent the WH as a seamless
connection between two such flat regions. From the well-behaved
embedded surface, we infer that the throat is free from any
geometrical irregularities. In the present framework, the flare-out
condition is satisfied without requiring energy conditions'
violation, which is different from GR. As established in the
preceding section, the matter content satisfies the energy bounds,
while the matter-geometry coupling in $f(R,T)$ gravity plays a major
role in supporting the WH structure. Figure \textbf{3} shows that
both WH models generate a well-defined embedding profile at the
throat. The rapid reduction of $z'(r)$ with increasing $r$ further
indicates that the geometry smoothly evolves from the throat toward
an asymptotically flatter spatial configuration, confirming the
expected WH embedding behavior for both models.
\begin{figure}[h!]
\centering
\epsfig{file=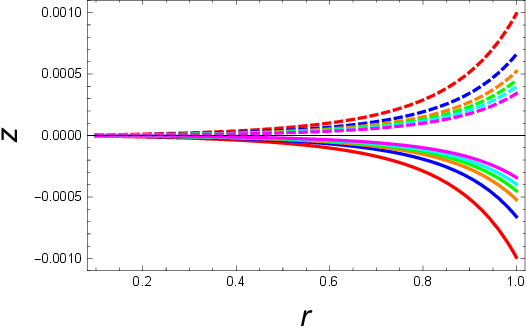,width=0.45\linewidth}\hfill\epsfig{file=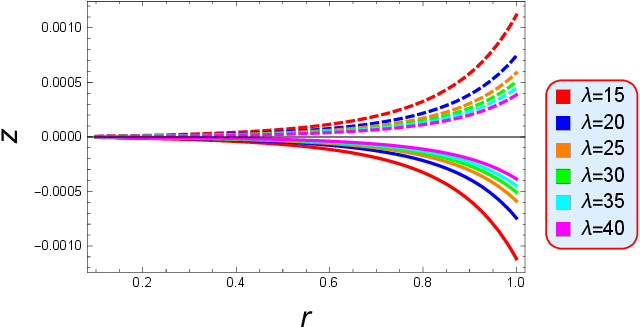,width=0.545\linewidth}
\epsfig{file=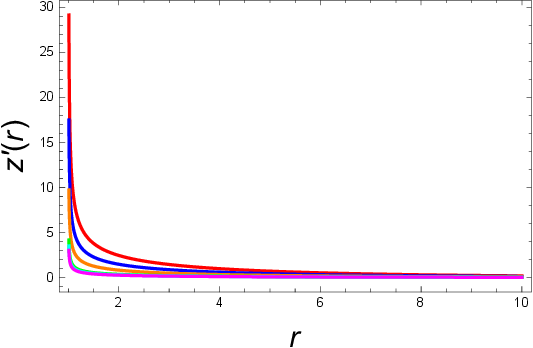,width=0.44\linewidth}\hfill\epsfig{file=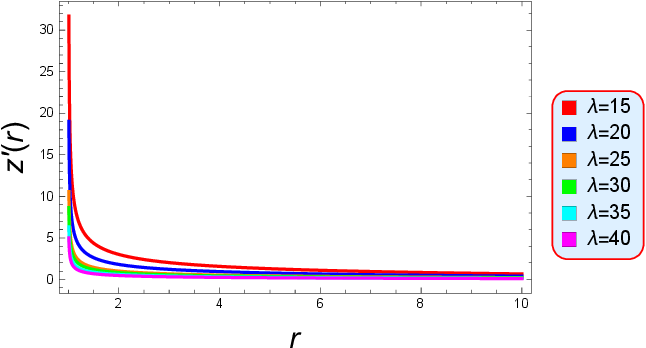,width=0.53\linewidth}
\caption{Embedding function and its derivative for model 1 (left)
and model 2 (right).} \label{fig:ShapeFunction}
\end{figure}

\subsection{Basic Stability Evaluation}

To make sure the WH solutions are physically viable, we need to see
how stable they are against small perturbations. The response of the
anisotropic matter supporting the WH to small perturbations offers a
useful stability assessment. For an anisotropic matter setup, we
define the radial and tangential propagation speeds as
\begin{equation}\label{36}
v_{r}^{2}=\frac{dp_{r}}{d\rho}, \quad
v_{t}^{2}=\frac{dp_{t}}{d\rho}.
\end{equation}
The requirement of causality imposes that both sound velocities
fulfill
\begin{equation}\label{37}
0\leq v_{r}^{2}\leq 1, \quad 0\leq v_{t}^{2}\leq 1,
\end{equation}
guaranteeing that all perturbations travel at speeds below the speed
of light \cite{bbb}. Figure \textbf{4} shows that both radial and
tangential sound speeds remain within the causal range for both
models. Thus, the propagation of perturbations remains subluminal,
indicating that the causality condition remains valid across the
radial domain. The parameter $\lambda$ changes the magnitude of
$v^{2}$, while both models remain causally well-behaved.
\begin{figure}[h!]
\centering
\epsfig{file=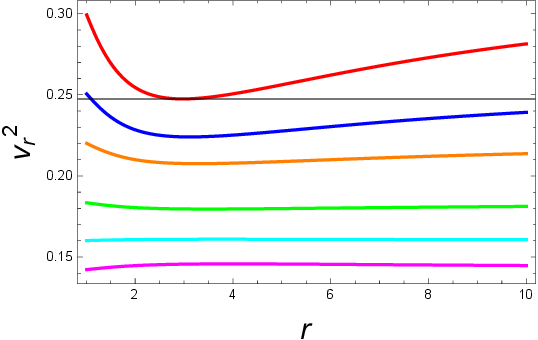,width=0.45\linewidth}\hfill\epsfig{file=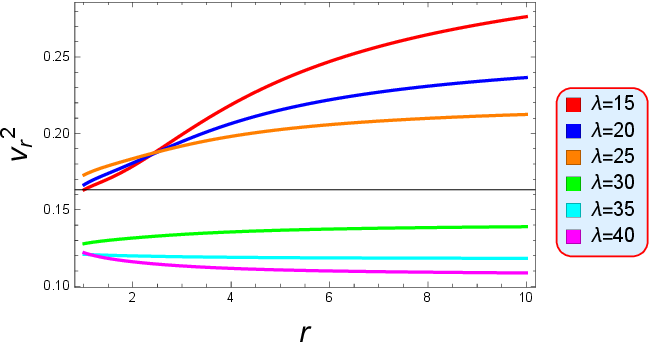,width=0.545\linewidth}
\epsfig{file=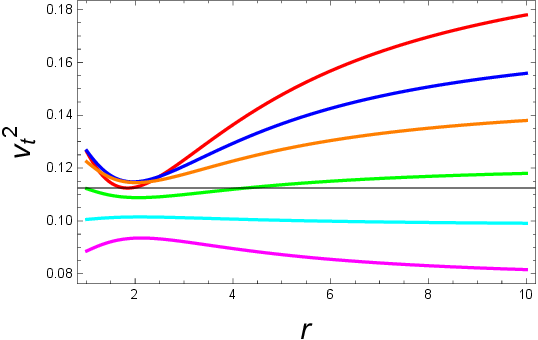,width=0.45\linewidth}\hfill\epsfig{file=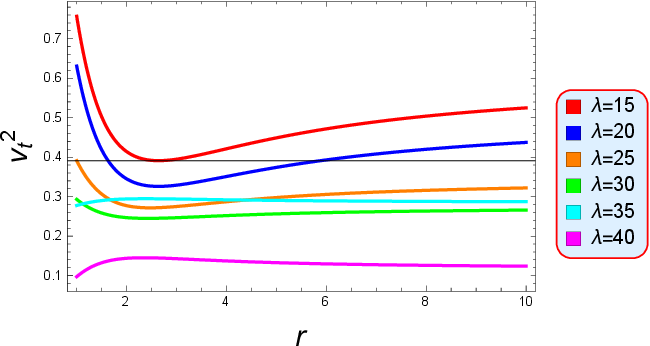,width=0.545\linewidth}
\epsfig{file=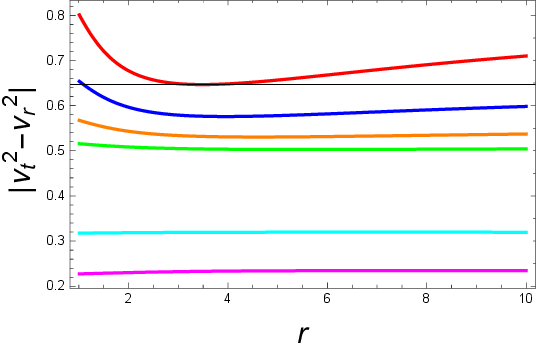,width=0.45\linewidth}\hfill\epsfig{file=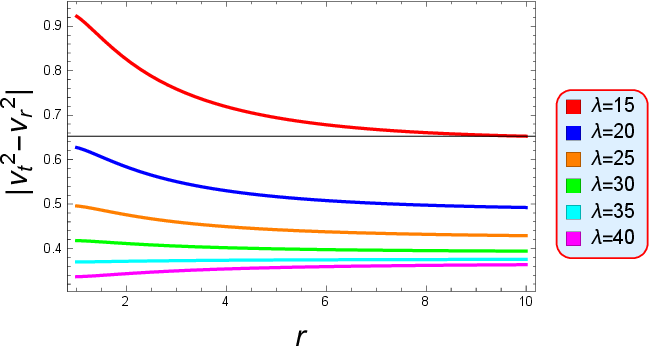,width=0.545\linewidth}
\caption{Causality and cracking conditions for model 1 (left) and
model 2 (right).} \label{fig:Causality}
\end{figure}

A further useful diagnostic for anisotropic matter is the cracking
criterion proposed by Herrera, under which the system is stable if
it obeys \cite{aaa}
\begin{equation}\label{38}
0\leq\left|v_{t}^{2}-v_{r}^{2}\right|\leq 1.
\end{equation}
Figure \textbf{4} (last row) shows that the above factor remains
below unity for both models, satisfying the cracking condition.
Hence, both configurations remain stable against the considered
cracking perturbations. Notably, the above conditions provide the
necessary criteria for assessing the physical viability and
stability of the proposed models. Given that no divergences occur
for $r>r_0$, the matter distribution maintains regularity over the
entire domain outside the central neck. The stability aspects
covered above ought to be understood as initial local checks of
physical soundness. To fully establish stability, one would need to
study dynamical perturbations of the geometric structure and matter
configuration, going outside the current effective treatment. Even
so, if we take the sound speed constraints together with the
cracking test results, they indicate that the rotating WH solutions
in $f(R,T)$ modified gravity setup are physically acceptable.

\section{Particle Dynamics and Rotational Effects}

Having established the rotating WH solutions, we turn our attention
to their dynamic behavior, specifically examining how angular
momentum influences the trajectories of test particles. The analysis
is built around frame dragging, the effective potential profile, and
the deviations that follow from the non-minimal matter-geometry
coupling. By examining both timelike and null geodesics, we gain
further perspective on the physical characteristics and potential
observability of this spacetime.

\subsection{Effective Geodesic Motion}

Test particle dynamics serve as a valuable probe of both the
physical nature and the observable features inherent to rotating
compact objects. The non-conservation of the EMT is a characteristic
feature of $f(R,T)$ gravity, leading to non-geodesic particle motion
due to matter-geometry coupling. However, before these additional
effects are included, we first examine the key geometry related
contributions under a standard geodesic approximation
\cite{ccc,eee}. This method is simple enough to solve and still
shows the main geometric effects. In this regime, test-particle
trajectories are determined by
\begin{equation}\label{39}
\frac{d^{2}x^{\mu}}{d\tau^{2}} + \Gamma^{\mu}_{\alpha\beta}
\frac{dx^{\alpha}}{d\tau} \frac{dx^{\beta}}{d\tau} =0,
\end{equation}
where $\tau$ is the affine parameter. We focus on trajectories lying
in the equatorial plane, $\theta=\pi/2$, for the metric in
Eq.\eqref{14}. As the metric is independent of $t$ and $\phi$, both
$E$ and $L$ remain unchanged, defined by
\begin{align}\label{40}
E&=-g_{tt}\dot{t}-g_{t\phi}\dot{\phi}, \\\label{41}
L&=g_{\phi\phi}\dot{\phi}+g_{t\phi}\dot{t}.
\end{align}
Employing the metric components, one arrives at
\begin{align}\label{42}
E&=N^2(r)\dot{t}+r^2\omega(r)\dot{\phi}, \\\label{43}
L&=r^2\dot{\phi}-r^2\omega(r)\dot{t}.
\end{align}
So, if we consider rotational effects only to linear order, assuming
a small angular velocity, we end up with
\begin{align}\label{44}
\dot{t} &= \frac{E-\omega(r)L}{N^2(r)}, \\\label{45} \dot{\phi} &=
\frac{L}{r^{2}} + \frac{\omega(r)E}{N^{2}(r)}.
\end{align}
The normalization condition gives us the radial equation of motion
\begin{equation}\label{46}
g_{\mu\nu}\dot{x}^{\mu}\dot{x}^{\nu} = -\gamma, \quad \gamma=
\begin{cases}
1, & \text{trajectories of massive particles},\\
0, & \text{trajectories of massless particles}.
\end{cases}
\end{equation}
Using Eqs.\eqref{44} and \eqref{45} in the metric relation
\eqref{46} leads to
\begin{equation}\label{47}
\frac{\dot{r}^{\,2}}{1-\dfrac{b(r)}{r}} =
\frac{E\left(E-2\omega(r)L\right)}{N^2(r)} -\frac{L^2}{r^2} -\gamma.
\end{equation}
So, Eq.\eqref{47} becomes
\begin{equation}\label{48}
\dot{r}^{\,2}+V_{\mathrm{eff}}(r)=0.
\end{equation}
Here, we introduce the effective potential as
\begin{equation}\label{49}
V_{\mathrm{eff}}(r) = -\left(1-\frac{b(r)}{r}\right) \left[
\frac{E\left(E-2\omega(r)L\right)}{N^2(r)} -\frac{L^2}{r^2} -\gamma
\right].
\end{equation}
The effects of both the WH spacetime and rotational frame dragging
are contained in Eq.\eqref{49}. More specifically, the presence of
the extra $\omega(r)EL$ term leads to modifications in the
trajectories compared with the non-rotating case. At the WH central
neck, the effective potential maintains a smooth and regular
behavior. So, physical bodies can pass through the WH as long as
$\dot{r}^{2} = 0$ at $r = r_0$. Figure \textbf{5} presents the
radial velocity profiles for both models. The profiles illustrate
how the WH geometry and rotation modify the radial motion of test
particles, with the effect being more pronounced near the throat and
gradually decreasing at larger $r$.
\begin{figure}[h!]
\centering
\epsfig{file=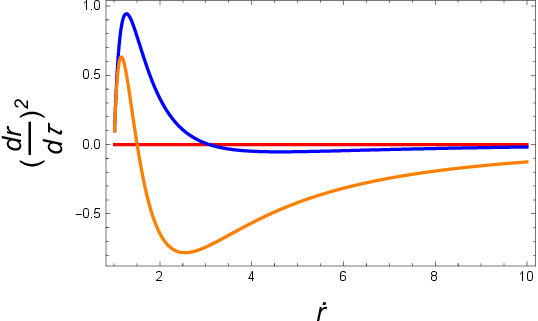,width=0.465\linewidth}\hfill\epsfig{file=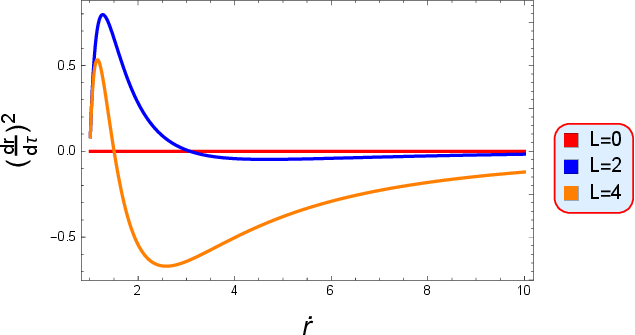,width=0.53\linewidth}
\caption{Radial velocity profile for model 1 (left) and model 2
(right).} \label{fig:Radial Velocity Profile}
\end{figure}

\subsection{Frame Dragging and Effective Potential}

Frame dragging shows up because of rotation, and it arises from the
off-diagonal metric term $g_{t\phi}$. Rotation makes inertial frames
follow the spacetime's motion. This shows up as changes in particle
trajectories and orbital structure. A zero angular-momentum observer
has $L = 0$, and its angular velocity is
\begin{equation}\label{50}
\Omega_{\mathrm{ZAMO}} = \frac{d\phi}{dt} = \omega(r).
\end{equation}
Even if a particle has no angular momentum, the dragging of inertial
frames still makes it rotate. As shown in Eq.\eqref{28}, $\omega(r)$
is obtained as $\frac{C}{r^{3}}$ in the large-$r$ limit, which is
known to be the typical Lense-Thirring behavior for slow rotators.
Thus, frame dragging is found to dominate close to the throat, while
it drops off rapidly at large distances, where the spacetime is seen
to revert to the standard flat form. An additional rotational effect
is observed in the profile of the effective potential \eqref{49}.
The presence of the $\omega(r)EL$ term relocates the stable and
unstable circular paths, modifies the return points, and deforms the
closed trajectories relative to the non-rotating limit. So, the
rotation of the spacetime leaves its mark on how matter particles
move around the WH. The orbital structure close to the throat is
particularly affected by the asymmetry introduced through frame
dragging, whereas standard motion is recovered at large $r$ because
$\omega(r)$ drops off. Frame dragging effects are found to be most
significant near the central neck, where the rotational profile is
at its strongest. Different effective gravitational forces are
experienced by co-rotating and counter-rotating trajectories, so the
orbital behavior is asymmetric.

Figure \textbf{6} shows that the effective potential strongly
depends on the angular momentum. Its zero points determine the
turning points, while its minima and maxima indicate stable and
unstable orbital regions, respectively. The similar behavior in both
models confirms the influence of rotation on particle dynamics.
Figure \textbf{7} depicts the rotational function $\omega(r)$
together with its corresponding frame dragging influence. The
profile follows the characteristic slow rotation behavior
$\omega(r)\sim C/r^3$, showing that the rotational contribution is
well-behaved and rapidly suppressed in the asymptotic region.
\begin{figure}[h!]
\centering
\epsfig{file=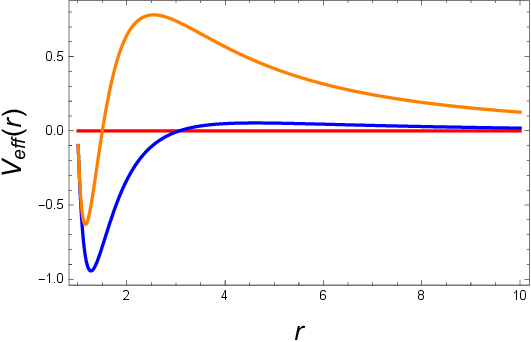,width=0.45\linewidth}\hfill\epsfig{file=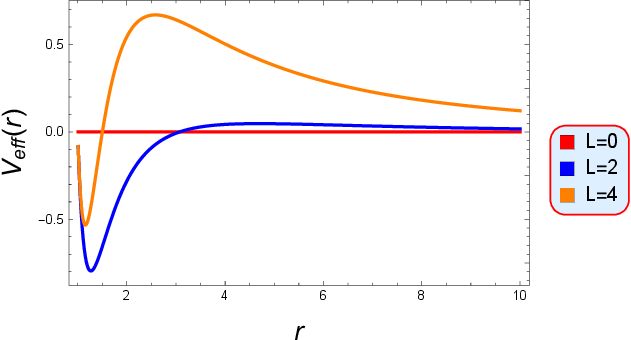,width=0.54\linewidth}
\caption{Effective potential for model 1 (left) and model 2
(right).} \label{fig:Effective Potential}
\end{figure}
\begin{figure}[h!]
\centering \epsfig{file=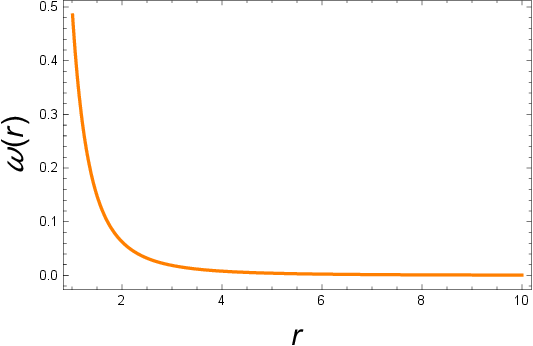,width=0.46\linewidth}
\caption{Rotation function $\omega(r)$ versus radial coordinate.}
\label{fig:Rotaion Function for model 1 and 2}
\end{figure}

\subsection{Non-Geodesic Behavior and Supplementary Force Contributions}

The effects of matter-geometry coupling in $f(R,T)$ gravity are
considered in the following analysis. As we have already discussed
that the EMT is generally not a conserved quantity, implying that
geodesic motion is no longer followed by massive test particles. The
revised equation of motion is provided by
\begin{equation}\label{51}
\frac{d^{2}x^{\mu}}{ds^{2}} + \Gamma^{\mu}_{\alpha\beta}
u^{\alpha}u^{\beta} = f^{\mu},
\end{equation}
where the extra force due to matter-geometry coupling is represented
by $f^{\mu}$. For a perfect fluid, this term is expressed as
\begin{equation}\label{52}
f^{\mu} = \frac{8\pi}{(\rho+p)(8\pi+f_{T})}
\left(g^{\mu\nu}-u^{\mu}u^{\nu}\right)\nabla_{\nu}p.
\end{equation}
When the pressure vanishes, the extra force term becomes zero,
restoring the usual geodesic motion of GR. Although the standard
relation is formulated for ideal fluids, introducing an effective
mean pressure allows it to be applied to non-isotropic matter
configurations. Since our matter source is anisotropic, an average
pressure is introduce as
\begin{equation}\label{53}
p=\frac{p_r+2p_t}{3}.
\end{equation}
Thus, the pressure gradient $p$ and combination of $\rho$ takes the
respective form
\begin{equation}\label{54}
\rho+p = \rho+\frac{p_r+2p_t}{3},
\end{equation}
and
\begin{equation}\label{55}
\frac{dp}{dr} = \frac{1}{3} \left( \frac{dp_r}{dr} +
2\frac{dp_t}{dr} \right),
\end{equation}
and, hence, the reformulated equation of motion \eqref{52} can be
cast as
\begin{equation}\label{56}
\frac{d^{2}x^{\mu}}{ds^{2}} +
\Gamma^{\mu}_{\alpha\beta}u^{\alpha}u^{\beta} =
\frac{8\pi}{(8\pi+\lambda)(\rho+p)} \left(g^{\mu
r}-u^{\mu}u^{r}\right) \frac{dp}{dr}.
\end{equation}
Finally, the extra force is obtained as
\begin{equation}\label{57}
f^{\mu} = \frac{8\pi} {(8\pi+\lambda)(\rho+p)} \left(g^{\mu
r}-u^{\mu}u^{r}\right) \frac{dp}{dr}.
\end{equation}
When considering $u^{r} = 0$, the above force is seen to have just a
radial part
\begin{align}\label{58}
f^{r}& = \frac{8\pi} {(8\pi+\lambda)(\rho+p)} g^{rr} \frac{dp}{dr},
\end{align}
which further becomes
\begin{align}\label{59}
f^{r}&= \frac{8\pi} {(8\pi+\lambda)(\rho+p)}
\left(1-\frac{b(r)}{r}\right) \frac{dp}{dr}.
\end{align}

The additional force modifies only the equations of motion for
massive test particles. Null rays feel no such influence; they trace
out geodesics set entirely by the spacetime curvature. Figure
\textbf{8} shows that the radial extra force is sensitive to the
shape function parameters $a$ and $\mu$, demonstrating how the WH
geometry and matter-geometry coupling jointly affect the radial
motion of massive particles. Figure \textbf{9} demonstrates that
interaction of matter with spacetime curvature in $f(R,T)$ gravity
reduces the effective potential barrier, making the WH more
traversable for massive particles, with the effect controlled by the
coupling parameter $\lambda$.
\begin{figure}[h!]
\centering
\epsfig{file=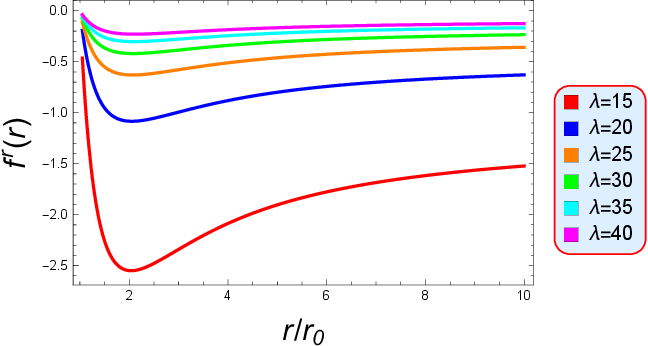,width=0.49\linewidth}\hfill\epsfig{file=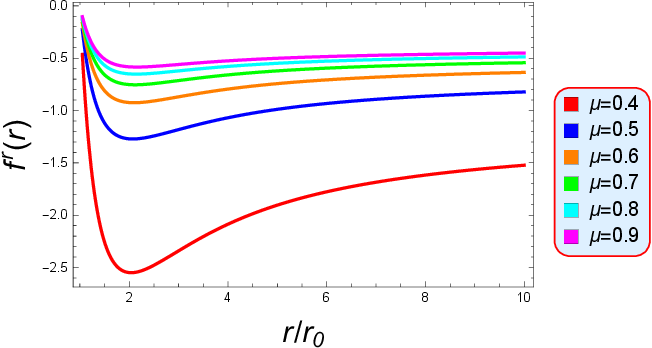,width=0.49\linewidth}
\epsfig{file=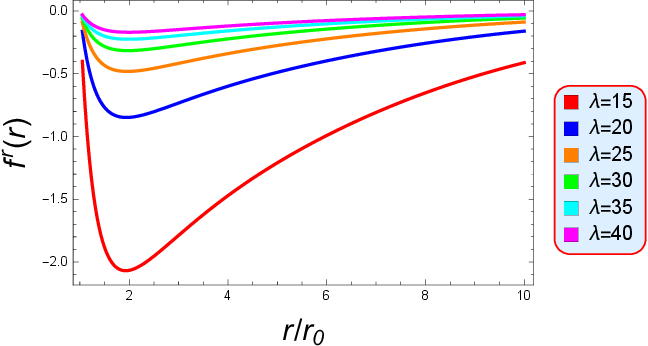,width=0.49\linewidth}\hfill\epsfig{file=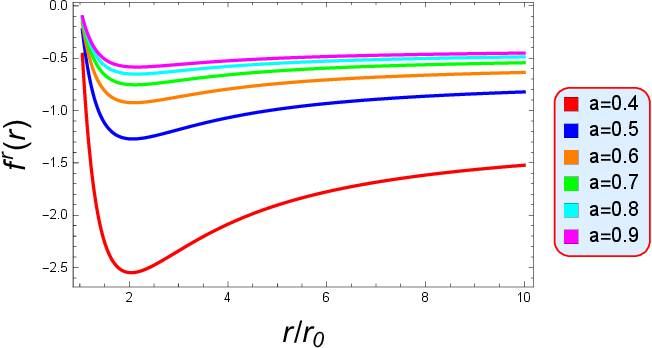,width=0.49\linewidth}
\caption{Radial extra force for model 1 (upper) and model 2
(lower).} \label{fig:Radial Extra Force}
\end{figure}
\begin{figure}[h!]
\centering
\epsfig{file=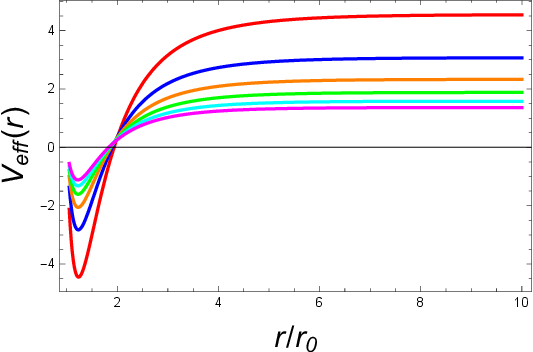,width=0.43\linewidth}\hfill\epsfig{file=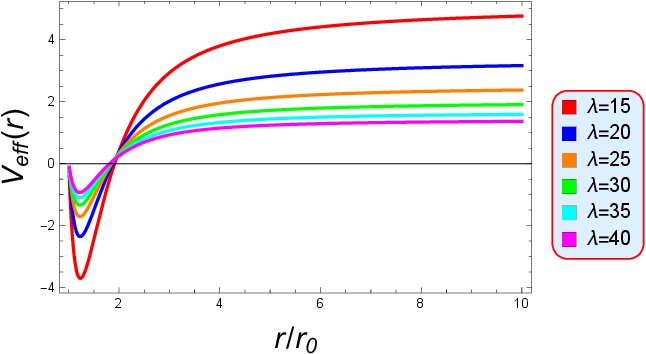,width=0.525\linewidth}
\epsfig{file=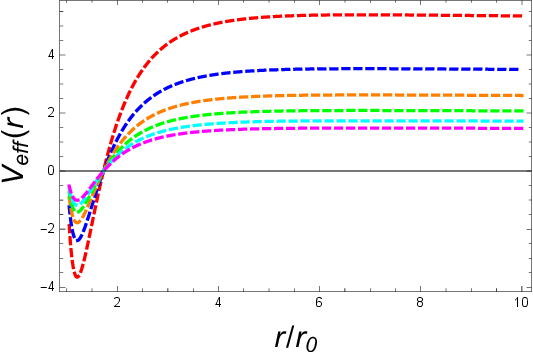,width=0.43\linewidth}\hfill\epsfig{file=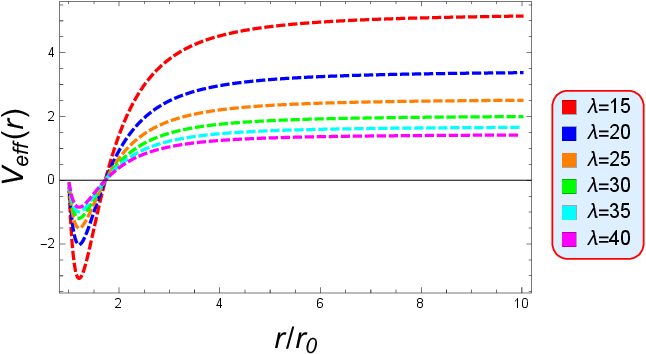,width=0.525\linewidth}
\caption{Effective potential for model 1 (left) and model 2 (right)
with (upper plots) and without (lower plots) extra force.}
\label{fig:Effective Potential with and without Extra Force}
\end{figure}

\subsection{Weak-field Lensing}

The WH geometry can also be investigated through gravitational
lensing, which serves as a key observational channel. Specifically,
the gravitational bending of light in the neighborhood of rotating
dense objects can manifest signs of frame dragging and the related
metric structure. Using the radial equation \eqref{47} for
$\gamma=0$ and introducing the impact parameter
\begin{equation}\label{60}
\xi_{c} = \frac{L}{E},
\end{equation}
the radial equation takes the form
\begin{equation}\label{61}
\frac{\dot{r}^2}{E^2\left(1 - \frac{b(r)}{r}\right)} = \frac{1 -
2\omega(r)\xi_c}{N^2(r)} -\frac{\xi_c^2}{r^2}.
\end{equation}

Using the radial equation of motion, the corresponding conditions
for circular photon trajectories are obtained as
\begin{equation}\label{62}
\mathcal{R}(r,\xi_c) = 0,\quad \frac{d\mathcal{R}(r,\xi_c)}{dr} = 0,
\end{equation}
where
\begin{equation}\label{63}
\mathcal{R}(r,\xi_c) = \frac{1 - 2\omega(r)\xi_c}{N^2(r)}
-\frac{\xi_c^2}{r^2}.
\end{equation}

Employing Eq.\eqref{62}, the bending angle of light takes the formal
expression
\begin{equation}\label{64}
\alpha (r_0) =
2\int_{r_0}^{\infty}\frac{dr}{\sqrt{\mathcal{R}(r,\xi_c)}} -\pi.
\end{equation}

The deflection angle is modified by the rotational function
$\omega(r)$ via frame dragging effects, thereby yielding different
bending behaviors for photons moving parallel and antiparallel to
the rotation. Further progress can be made by working in the
weak-field approximation, where the closest approach far exceeds the
throat radius, i.e., $r_0>>r_{th}$. With the effects of $b(r)$ and
$\omega(r)$ becoming negligible, the radial function can be written
as
\begin{equation}\label{65}
\mathcal{R}(r,\xi_c)\approx \frac{1}{N^2(r)}(1 - 2\omega (r)\xi_c) -
\frac{\xi_c^2}{r^2}.
\end{equation}
Employing this expansion in Eq.\eqref{64}, we obtain the deflection
angle in schematic form as
\begin{equation}\label{66}
\alpha (r_0)\approx \alpha_{\mathrm{static}}(r_0) + \delta
\alpha_{\mathrm{rot}}(r_0).
\end{equation}
In this expression, $\alpha_{static}$ refers to the static WH
contribution, and $\delta\alpha_{rot}$ is the leading-order
rotational term. When rotation is turned off, the expression reduces
to
\begin{equation}\label{67}
\alpha_{\mathrm{static}}(r_0) =
2\int_{r_0}^{\infty}\frac{dr}{\sqrt{\frac{1}{N^2(r)}} -
\frac{\xi_c^2}{r^2}} -\pi.
\end{equation}
\begin{figure}[h!]
\centering
\epsfig{file=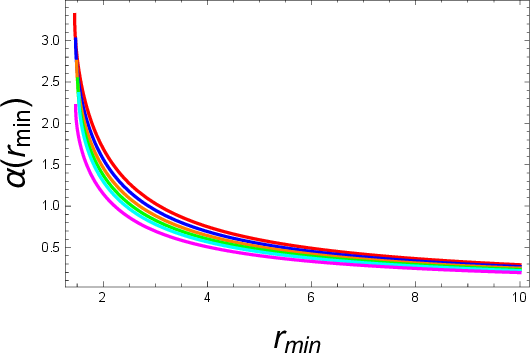,width=0.43\linewidth}\hfill\epsfig{file=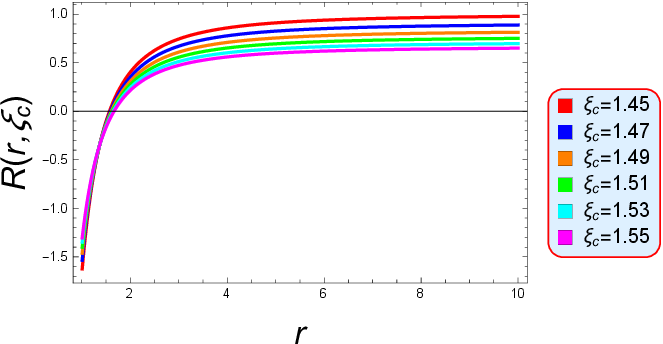,width=0.545\linewidth}
\caption{Deflection angle (left) and radial lensing function
(right).} \label{fig:Deflection angle and Radial lensing function}
\end{figure}

Figure \textbf{10} illustrates the gravitational lensing in the
rotating WH spacetime. The left panel shows that light bending is
strongest for photons passing closest to the throat, with deflection
decreasing monotonically to zero in the asymptotic region. The right
panel confirms the existence of allowed and forbidden photon
regions, with the critical impact parameter marking the shadow
boundary. These results demonstrate that the WH geometry produces
realistic lensing signatures consistent with compact objects.

\section{Conclusions and Future Directions}

Traversable WHs rank among the most captivating outcomes obtained
from the gravitational field equations. Such configurations provide
the spacetime fabric with a distinct topological structure that
connects two distant areas via a slender passage. Aside from
geometric considerations, these entities are beneficial for
analyzing regimes of strong gravity, dense astrophysical objects,
and prospective modifications to GR. Wormhole geometries have
attracted increasing attention because of their interesting
structure and physical properties. In this study, rotating
traversable WH configurations are examined in $f(R,T)$ gravity for
two different forms of the shape function. We began by formulating
$f(R,T)$ gravity and deriving the field equations for a linear
model. Subsequently, we made use of the rotating WH geometry to
evaluate the physical quantities and to study the geometric as well
as physical behavior of the solutions with a particular focus on the
linear functional form such as $f(R,T)$= $R$+$\lambda T$. The energy
conditions for each WH model have been studied across a variety of
parameter values. Furthermore, we thoroughly examined the embedding
diagrams, preliminary stability analysis, effective geodesic motion,
frame dragging, effective potential, non-geodesic motion due to
extra-force contribution, weak-field lensing signatures and found
significant findings. Our analysis used the parameters
$\lambda=15,20,25,30,35,40$, $a=0.4,0.5,0.6,0.7,0.8,0.9$,
$\mu=0.4,0.5,0.6,0.7,0.8,0.9$, and $L=0,2,4$. The significant
results obtained from this study are highlighted below.
\begin{itemize}
\item Both the shape functions satisfy the basic viability conditions.
It is found that both geometries are regular, asymptotically flat,
and possess a well-defined throat (Figure \textbf{1}).

\item We examine the energy conditions for multiple values of
$\lambda$. These conditions remain positive throughout the
considered region (Figure \textbf{2}). Their satisfaction at the
throat indicates that non-standard matter is not necessary to
sustain the WH geometry, setting these models apart from
conventional GR WHs and several modified gravity solutions. The
variations in $\lambda$ also modify the profiles of these
conditions, demonstrating its role in the physical behavior and
stability of the WH configurations.

\item The embedding diagrams presents both
positive and negative sectors (Figure \textbf{3}). These results
confirm the internal consistency of our WH models.

\item Both sound speeds remain between
$0$ and $1$ in both models (Figure \textbf{4}). This indicates that
the models satisfy the causality condition and allow subluminal
propagation for all considered values of $\lambda$. The condition
$|v_{r}^2-v_{t}^2|<1$ is also fulfilled in both models, which
demonstrates their stability against cracking perturbations.

\item The radial velocity profile for
$\lambda$ in both models indicates that the effect of the WH
geometry and rotation is stronger near the throat and decreases with
increasing $r$ (Figure \textbf{5}).

\item  It is shown how angular momentum shapes
the effective potential, with its zeros and extrema defining the
possible turning points and orbital stability regions (Figure
\textbf{6}).

\item The behavior of $\omega(r)$ confirms that frame dragging is strongest
close to the throat and rapidly decreases in the asymptotic domains
(Figure \textbf{7}).

\item The radial extra force
for both models shows that around the throat, the force is negative
and reaches its minimum around $r/r_0\approx2$, after which it
gradually returns to zero (Figure \textbf{8}). The magnitude of this
force decreases as the parameters $a$, $\mu$, and $\lambda$
increase, implying a weaker effect on radial particle motion.

\item Increasing the matter-geometry interaction can lower the effective
potential barrier, thereby supporting easier traversal of the WH by
massive particles (Figure \textbf{9}).

\item The weak-field lensing behavior
of the rotating WHs (Figure \textbf{10}). Deflection shows its
strongest behavior near the throat and gradually decreases with
distance. The critical impact parameter $\xi_c$ determine the
boundary of the photon shadow by separating the allowed and
forbidden photon regions.
\end{itemize}

Overall, the analysis establishes rotating WHs as viable stellar
configurations with a broad range of physical features. Their
regular traversable structure, energy conditions' behavior,
rotational dynamics, particle motion, and lensing related signatures
arise within the matter-geometry coupled framework. Table \textbf{1}
provides a direct comparison with GR and existing modified gravity
WH models, emphasizing the distinctive aspects of the present study.
\begin{table}[h!]
\centering \tiny \setlength{\tabcolsep}{3.2pt}
\renewcommand{\arraystretch}{3.29}
\caption {Comparative analysis of WH models across various
gravitational frameworks.} \label{tab:wh_comparison}
\begin{tabular}{|p{2.7cm}|p{1.65cm}|p{1.3cm}|p{1.65cm}|p{1.55cm}|p{1.55cm}|p{2.5cm}|}
\hline \textbf{Source} & \textbf{Gravitational Frameworks} &
\textbf{Rotation} & \textbf{Energy Constraints} & \textbf{Matter
Distribution} &
\textbf{Geodesic} & \textbf{Key Characteristics} \\
\hline Morris \& Thorne  \cite{22aa} & GR & Absent & Violated &
Exotic matter & Not included & Pioneering traversable WH proposal \\
\hline Teo  \cite{22bb} & GR & Present & Violated &
Exotic matter & Restricted & Rotating WH featuring frame dragging effect \\
\hline Lobo  \cite{22cc} & GR & Absent & Violated &
Exotic matter & Not included & Static WH with matter sources that violate standard energy conditions \\
\hline Zubair et al.  \cite{22dd} & $f(R,T)$ & Absent & May hold &
Non-isotropic fluid distribution & Not included & Impacts of matter-geometry coupling \\
\hline Sharif \& Waseem  \cite{22ee} & $f(R,T)$ & Absent & Partially
satisfied &
Anisotropic fluid & Not included & Static anisotropic WH solutions \\
\hline Filho et al. \cite{22gg} & $f(R,T)$ & Absent & dependent on
the model &
Non-isotropic fluid distribution & Included & Potential observable features and lensing effects \\
\hline \textbf{Current Study} & $f(R,T)=R+\lambda T$ (Slow rotation)
& Present & Energy conditions hold & Anisotropic fluid & Included &
Rotating traversable WH free from exotic matter, encompasses stability, geodesics, and lensing \\
\hline
\end{tabular}
\end{table}

Future work can study higher order rotation effects, non-equatorial
photon paths, and numerical analysis of shadows and lensing.
Quasinormal modes, accretion, and gravitational wave signals from
rotating WHs are also important topics for further study.
\\\\
\textbf{Data Availability:} Data is provided within the manuscript.

\end{document}